\documentclass[9pt,twocolumn,twoside]{osajnl}

\usepackage{tikz}
\newcommand*\circled[1]{\tikz[baseline=(char.base)]{
    \node[shape=circle, draw, inner sep=1pt, 
        minimum height=10pt] (char) {#1};}}

\journal{ol} 

\setboolean{shortarticle}{true}

\title{Loss compensation symmetry of unequally sized dielectric cylinders with gain and loss}

\author[1,2]{Anton V. Hlushchenko}
\author[2]{Vitalii I. Shcherbinin}
\author[3]{Denis V. Novitsky}
\author[1,*]{Vladimir R. Tuz}

\affil[1]{State Key Laboratory of Integrated Optoelectronics, College of Electronic Science and Engineering, International Center of Future Science, Jilin University, 2699 Qianjin Street, Changchun, 130012, China}
\affil[2]{National Science Center “Kharkiv Institute of Physics and Technology”, National Academy of Sciences of Ukraine, 1 Akademicheskaya Street, Kharkiv, 61108, Ukraine}
\affil[3]{B. I. Stepanov Institute of Physics, National Academy of Sciences of Belarus, 68 Nezavisimosti Avenue, Minsk 220072, Belarus}
\affil[*]{Corresponding author: tvr@jlu.edu.cn}

\begin{abstract}
A rigorous analytical approach is applied to solve the eigenvalue problem for a pair of circular dielectric cylinders with complex permittivity. This approach relies on field expansion in terms of two sets of orthogonal azimuthal modes, which are coupled due to finite distance between the cylinders. We investigate the ability of a gain-dielectric cylinder operated in the fundamental TM mode to compensate material losses of a larger cylinder operated in the higher-order radial TM mode. To achieve such a loss compensation phenomenon, a simple design strategy is developed. It is shown that this phenomenon can be achieved for a certain distance between the cylinders, which is associated with the exceptional point of the system. For smaller distances, the adverse impact of high-order azimuthal (hybrid) modes are found to be essential. The results obtained are validated against full-wave simulations.
\end{abstract}

\setboolean{displaycopyright}{true}

\begin{document}

\maketitle

Active nanophotonics is multidisciplinary research at a meeting ground of applied physics, optics, material science, and engineering \cite{Krasnok_IEEE_2020}. It combines the latest advances in nanotechnology with gain materials and appears as a platform for the optical implementation of various concepts in non-Hermitian physics, including parity-time symmetric ($\mathcal{PT}$-symmetric) systems and exceptional points (EPs) \cite{Zyablovsky_2014, Feng_NatPhotonics_2017, Ozdemir_NatMat_2019, Miri_Science_2019}. In optics, $\mathcal{PT}$-symmetry corresponds to balanced  gain and loss distribution over the system. Such a gain-loss balance gives rise to EPs which are spectral singularities with coalesced eigenvalues and eigenvectors of the operator on the state space of the system \cite{Heiss_PhysRevE_2000, Heiss_JPhysA_2012}. EPs are subject of much current interest owing to their unusual properties, which enable intriguing phenomena and applications in non-Hermitian photonic systems. In particular, EPs are used as building blocks for designing polarizers \cite{Hassan_2017}, coherent perfect absorbers and lasers \cite{Longhi_2010,Wong_2016}, sensors \cite{Chen_2017,Hodaei_2017} and laser gyroscopes \cite{Hokmabadi_2019,Lai_2019} with enhanced sensitivity, slow-light structures \cite{Goldzak_2018}, and others.

Among different systems, optical directional couplers are most probably the simplest devices capable of exhibiting $\mathcal{PT}$-symmetry \cite{Guo_PhysRevLett_2009, Ruter_NatPhys_2010}. They are in the form of two coupled waveguides with balanced gain and loss. When these two waveguides are weakly coupled, and the gain and loss are small enough to behave like perturbations, the approximate coupled mode theory can be applied to describe interactions of the waveguide modes \cite{El-Ganainy_OptLett_2007}. This theory can also be extended to weakly nonlinear optical couplers with balanced gain and loss \cite{Chen_IEEE_1992, Sukhorukov_PhysRevA_2010, Alexeeva_PhysRevA_2012}.

Despite the benefits from the $\mathcal{PT}$-symmetrical directional couplers, their realization in nanophotonics is subject to strict design constraints on the waveguide parameters. To implement a more flexible design solution, dissimilar waveguide couplers can be used \cite{Walasik_NewJPhys_2017}. In such a device, the geometrical or material parameters of the waveguides forming the coupler are different. Although in dissimilar couplers the gain and loss distribution does not fulfill the $\mathcal{PT}$-symmetry conditions, they can support modes with real propagation constants and provide stable mode propagation and energy conservation, when the gain and loss characteristics in use are properly selected. Another feature, which distinguishes the dissimilar couplers from conventional $\mathcal{PT}$-symmetrical ones, is the ability to achieve a real propagation constant for only the desired mode of the coupler by selecting a unique pair of gain and loss values. Below these values, the desired mode has a negligible attenuation. For gain and loss above these values, this mode experiences gain properties, similar to those inherent in the $\mathcal{PT}$-symmetric systems. In the nonlinear regime, the dissimilar couplers acquire bi- and multistability states making the device more versatile for use in all-optical signal processing, ultrafast switching, and memory applications \cite{Walasik_OptLett_2015, Govindarajan_OptLett_2019, Govindarajan_OptLett_2020}. It should be noted that, for the dissimilar waveguide couplers, the weakly coupled theory may be inadequate and a more rigorous analysis of dielectric waveguide structures is required \cite{Thompson_Lightwave_1986}.

Dissimilar waveguide couplers belong to a general class of non-$\mathcal{PT}$-symmetric structures with unbalanced gain and loss \cite{Ghosh_SciRep_2016, Novitsky_PhysRevB_2017, Huang_OptExp_2017, Huang_OptExp_2019, Abdrabou_JOSAB_2019}. Their properties can be explained by the fact that EPs can exist in a large family of non-Hermitian systems, which do not necessarily satisfy the $\mathcal{PT}$-symmetry condition. In particular, the underlying formation mechanism of the EPs in dissimilar couplers is associated with the interaction between distinct propagating modes supported by the waveguides. Among possible propagation conditions for these modes, there is a regime, in which losses presented in one part of the system are totally compensated by gain introduced in another part. Such a regime can be referred to as loss compensation symmetry (LC-symmetry) \cite{Klimov_LaserPhysLett_2018}. The determination of loss compensation conditions is essential for nanophotonics and metamaterial research \cite{Krasnok_IEEE_2020, Hess_2012}.

In this Letter, we apply a rigorous theoretical approach to study dissimilar waveguide coupler operating in the LC-symmetric multi-mode conditions. The approach is based on the mode-matching technique and expansion of the waveguide fields in terms of  cylindrical harmonics \cite{White_JOSAB_2002, Faouri_ElMag_2015}. Such an approach does not suffer from the restrictions on the gain, loss, and mode coupling in the system, and can be extended to couplers with time-varying medium \cite{nerukh_2012}. We show that the EP appears under the crossing condition for the dispersion characteristics of two coupled modes with different order and the same type, and manifests itself in full loss compensation in the coupler. Under this condition, coupled modes somewhat resemble those of the $\mathcal{PT}$-symmetric systems.

\begin{figure}[t!]
\centering
\fbox{\includegraphics[width=\linewidth]{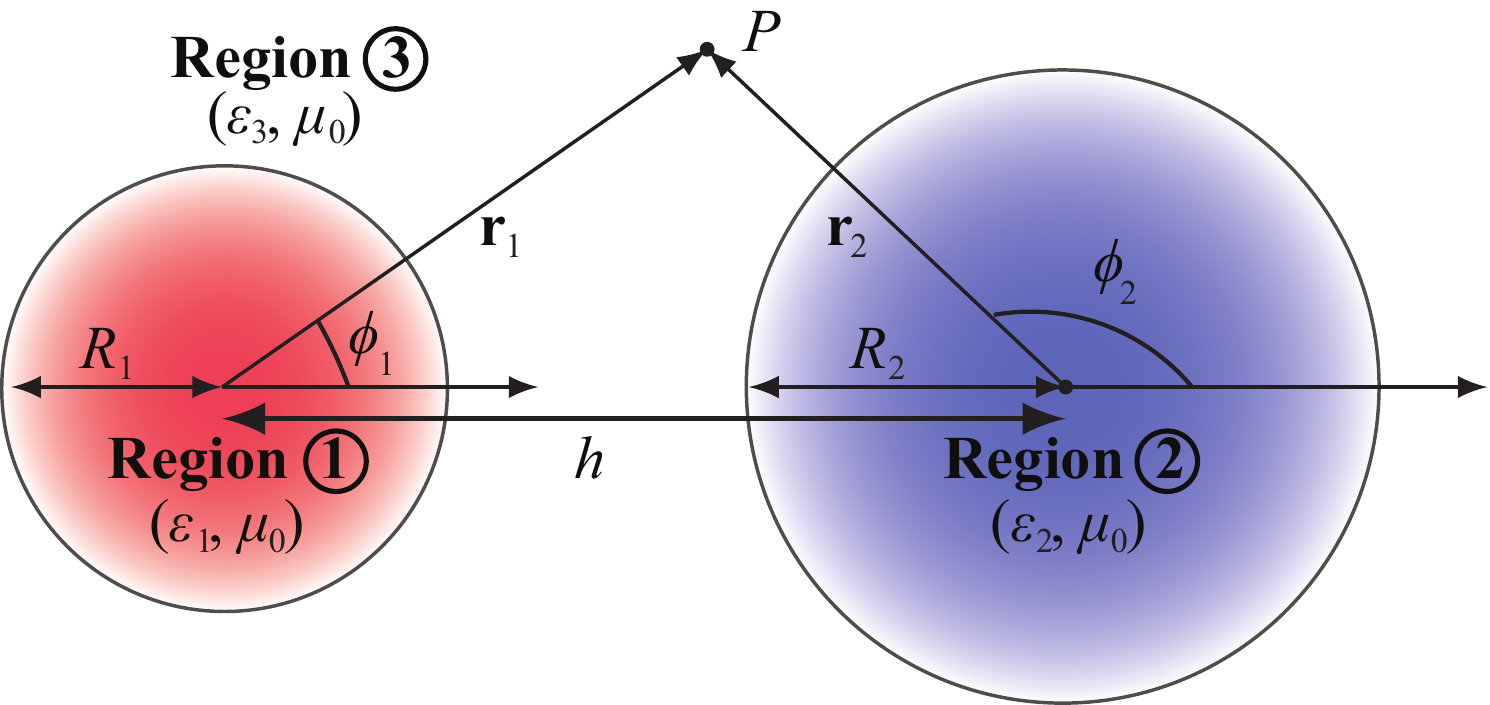}}
\caption{A pair of dielectric cylinders immersed in an infinite medium.}
\label{fig:cylinders}
\end{figure}

Consider a pair of nonoverlapping dielectric cylinders of the radii $R_1$ and $R_2$ (Fig. \ref{fig:cylinders}). Our prime interest is in the eigenfrequencies and eigenfields of the coupled cylinders immersed in an infinite medium. To solve the eigenvalue problem of interest, we consider three regions, each having the permeability $\mu_0$. Regions \circled{\small{1}} and \circled{\small{2}} are the transverse cross-sections of the first and second cylinder with the permittivity $\varepsilon_1$ and $\varepsilon_2$, respectively. The ambient space presents Region \circled{\small{3}} having the permittivity $\varepsilon_3$. For definiteness, in the following, we assume that the constitutive parameters under consideration satisfy the following conditions: Im$(\varepsilon_1)<0$ (gain material), Im$(\varepsilon_2)>0$ (lossy material), $\varepsilon_3 \in\Re$, Re$(\varepsilon_1)>\varepsilon_3$ and Re$(\varepsilon_2)>\varepsilon_3$.

Let us introduce two polar coordinate systems $(r_1,\phi_1)$ and $(r_2,\phi_2)$ related to both cylinders and write the following solutions of the eigenvalue problem for Region \circled{\small{1}} 
\begin{equation}\label{field_1}
\begin{split}
    E_z^1=\sum_{n=-N}^{N}{A_n^1J_n(k_{p,1}r_1)e^{in\phi_1}}, ~~ H_z^1=\sum_{n=-N}^{N}{B_n^1J_n(k_{p,1}r_1)e^{in\phi_1}},
\end{split}
\end{equation}
Region \circled{\small{2}} 
\begin{equation}\label{field_2}
\begin{split}
    E_z^2=\sum_{n=-N}^{N}{A_n^2J_n(k_{p,2}r_2)e^{in\phi_2}},~~ H_z^2=\sum_{n=-N}^{N}{B_n^2J_n(k_{p,2}r_2)e^{in\phi_2}},
\end{split}
\end{equation}
and Region \circled{\small{3}} 
\begin{equation}\label{field_3}
\begin{split}
    &E_z^3=\sum_{n=-N}^{N}{C_n^{1}H_n^{(1,2)}(k_{p,3}r_1)e^{in\phi_1}}+\sum_{n=-N}^{N}{C_n^2H^{(1,2)}_n(k_{p,3}r_2)e^{in\phi_2}},\\
    &H_z^3=\sum_{n=-N}^{N}{D_n^1H^{(1,2)}_n(k_{p,3}r_1)e^{in\phi_1}}+\sum_{n=-N}^{N}{D_n^2H^{(1,2)}_n(k_{p,3}r_2)e^{in\phi_2}}.
\end{split}
\end{equation}
where $\{A_n^1, A_n^2, B_n^1, B_n^2, C_n^1, C_n^2, D_n^1, D_n^2\}$ are the unknown amplitudes of azimuthal harmonics, $k_{p,i}^2=k^2_i-k_z^2$, $k_i=k_0\varepsilon_{r_i}$  $(i=1,2,3)$, $\varepsilon_{r_i}=\varepsilon_i/\varepsilon_0$ is the relative permittivity, $k_0^2=\omega^2\varepsilon_0\mu_0$, $J_n(\cdot)$ is the Bessel function, $H_n^{(1)}(\cdot)$ and $H_n^{(2)}(\cdot)$ are the Hankel functions of the first and second kind, respectively, field factor of the form $\exp{[-i(\omega t-k_zz)]}$ is assumed and omitted. Note that the transverse field components $(E_r, H_r, E_{\phi}, H_{\phi} )$ in all regions can be readily expressed in terms of axial components Eqs.~(\ref{field_1})-(\ref{field_3}) from the Maxwell equations.

In Region \circled{\small{3}}, all field components must decay with a distance from the cylinders and satisfy the radiation boundary condition at infinity. Since the transverse wavenumber $k_{p,3}$ has generally a complex value for cylinders with gain and loss, one needs to take care of proper kind of the Hankel function in \eqref{field_3} to fulfill this condition. The Hankel function of the first and second kind must be used, if Im${(k_{p,3})}>0$ and Im${(k_{p,3})}<0$, respectively.

The fields Eqs.~(\ref{field_1})-(\ref{field_3}) are subject to the continuity conditions at the interfaces between the regions \circled{\small{1}}-\circled{\small{3}}:
\begin{align}\label{boun_cond}
&E_z^1(R_1,\phi_1)=E_z^3(R_1,\phi_1), & H_z^1(R_1,\phi_1)=H_z^3(R_1,\phi_1), \notag \\
&E_{\phi}^1(R_1,\phi_1)=E_{\phi}^3(R_1,\phi_1), & H_{\phi}^1(R_1,\phi_1)=H_{\phi}^3(R_1,\phi_1), \\
&E_z^2(R_2,\phi_2)=E_z^3(R_2,\phi_2), & H_z^2(R_2,\phi_2)=H_z^3(R_2,\phi_2), \notag \\
&E_{\phi}^2(R_2,\phi_2)=E_{\phi}^3(R_2,\phi_2), & H_{\phi}^2(R_2,\phi_2)=H_{\phi}^3(R_2,\phi_2). \notag
\end{align}

To implement the conditions \eqref{boun_cond}, one needs to express the field of the ambient space (Region \circled{\small{3}}) in terms of the coordinates $(r_1,\phi_1)$ or $(r_2,\phi_2)$. This can be done with the use of the Graf addition theorem \cite{Abramowitz_1965}:
\begin{equation}\label{graf_th}
\begin{split}
B_n(r_1)e^{\pm in\phi_1}=\sum_{k=-N}^{N}{B_{n+k}(h)J_k(r_2)e^{\mp i k\phi_2}}e ^{\pm i k\pi}, \\
B_n(r_2)e^{\pm in\phi_2}=\sum_{k=-N}^{N}{B_{n+k}(h)J_k(r_1)e^{\mp i k\phi_1}}e ^{\pm i n\pi},
\end{split}
\end{equation}
where $B_n(\cdot)$ is the $n$-th order cylindrical function and $h$ is the distance between the centers of two nonoverlapping cylinders (Fig. \ref{fig:cylinders}).

Substituting the fields Eqs.~(\ref{field_1})-(\ref{field_3}) into \eqref{boun_cond}, and using \eqref{graf_th} and orthogonal properties of the basis modes, one obtains the system of equations for the unknown amplitudes of the azimuthal harmonics. The system of equations has non-trivial solutions, once its determinant is zero. This condition yields the desired dispersion relation for a pair of dielectric cylinders immersed in ambient space.

It is our prime concern to investigate the possibility of LC-symmetry in unequally sized dielectric cylinder with gain and loss. At a single frequency the cylinders are assumed to operate in the TM$_{0m}$-TM$_{0n}$ mode pair ($n\neq m$). The loss compensation can be achieved for a certain distance between the cylinders. Such a distance is known as a LC-symmetry threshold and satisfies the following conditions \cite{Klimov_LaserPhysLett_2018}: $\textrm{Re}{(k_{z,0m})}=\textrm{Re}{(k_{z,0n})}$ and $\textrm{Im}{(k_{z,0m})}=\textrm{Im}{(k_{z,0n})}=0$. The first condition mainly determines the operating frequency. This frequency corresponds to the intersection of the dispersion curves for the TM$_{0m}$ and TM$_{0n}$ modes. To achieve such an intersection one may first neglect gain, loss, and mode coupling ($h=\infty$), and change the dimensions of either or both dielectric cylinders. For $R_1=10$ [$\mu$m], $R_2=19$ [$\mu$m] and $\varepsilon_{r_1}=\varepsilon_{r_2}=12$, $\varepsilon_{r_3}=1$ such a design procedure yields the operating frequency $f=5.4961$ [THz] for the TM$_{01}$-TM$_{02}$ mode pair. At this frequency crossing or anti-crossing of the dispersion curves are observed, depending on the coupling between cylinders \cite{Novotny_AmJPhys_2010, Liu_Optica_2014, Tuz_Superlatt_2017}.

\begin{figure}[t!]
\centering
\fbox{\includegraphics[width=\linewidth]{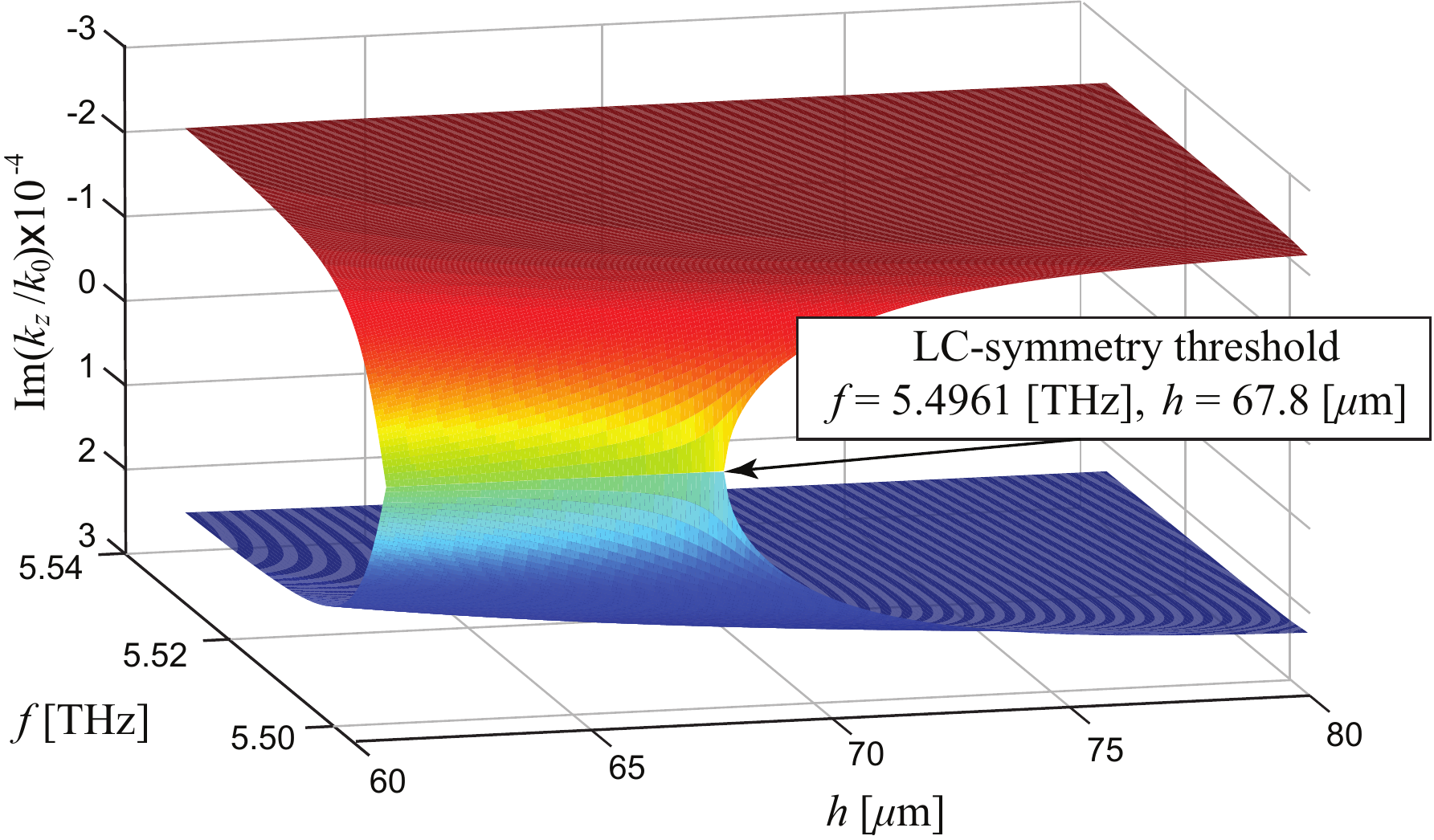}}
\caption{LC-symmetry threshold for the TM$_{01}$ and TM$_{02}$ modes of coupled dielectric cylinders with the radii $R_1=10$ [$\mu$m] and $R_2=19$ [$\mu$m],  and the relative permittivity  $\varepsilon_{r_1}=12(1-i1\times10^{-4})$ and $\varepsilon_{r_2}=12(1+i0.776\times10^{-4})$.}
\label{fig:surf}
\end{figure}

Next we take into account gain and loss properties of the dielectric cylinders and vary the distance $h$ between them. In this case, the eigenvalues of TM$_{0m}$ and TM$_{0n}$ modes may coalesce at a certain $h$ and thus an EP arises \cite{Heiss_JPhysA_2012, Ozdemir_NatMat_2019}. At the same time, to achieve real-valued eigenvalues under the LC-symmetry threshold one needs to keep the imaginary part of the permittivity $\varepsilon_{r_1}$ and $\varepsilon_{r_2}$ relatively small and to vary this part for either of two cylinders. We set $\varepsilon_{r_1}$ equal to $12(1-i1\times10^{-4})$. In this case, LC-symmetry threshold of the TM$_{01}$-TM$_{02}$ mode pair is observed for $h=67.8$ [$\mu$m], $\varepsilon_{r_2}=12(1+i0.776\times10^{-4})$, and $k_z=0.1795 $ [rad/$\mu$m] (Fig. \ref{fig:surf}). 

\begin{figure}[h]
\centering
\fbox{\includegraphics[width=\linewidth]{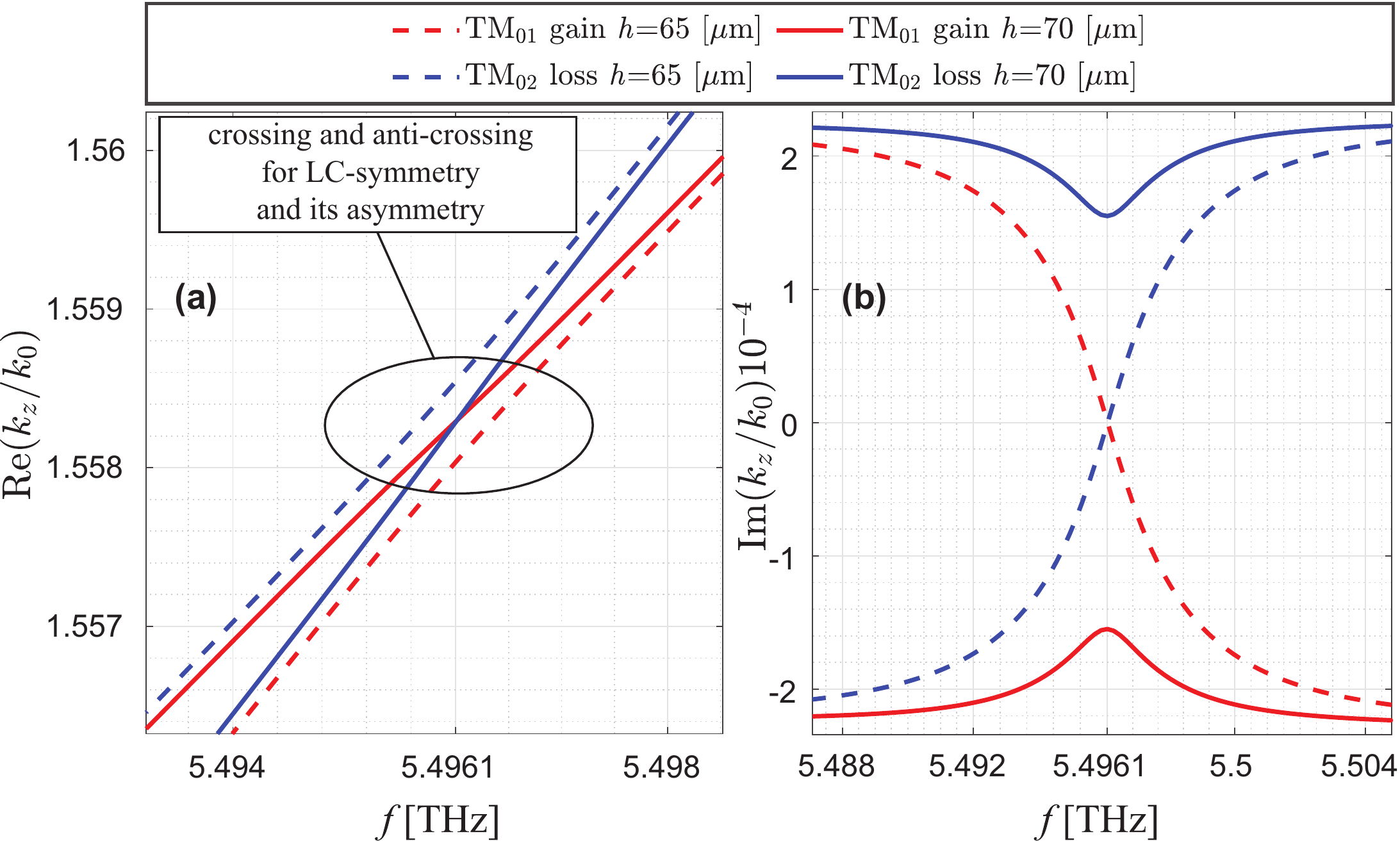}}
\caption{Crossing and anti-crossing of the dispersion curves of the TM$_{01}$ and TM$_{02}$ modes above ($h=70$ [$\mu$m]) and bellow ($h=65$ [$\mu$m]) the LC-symmetry threshold, respectively.}
\label{fig:cross}
\end{figure}

Above and below the threshold the dispersion curves of the TM$_{01}$ and TM$_{02}$ modes exhibit crossing and anti-crossing behavior, respectively (Fig. \ref{fig:cross}). In general, the gain and loss properties of the cylinders have an additional effect on the operating frequency. However, this effect is negligible for the low-loss and low-gain dielectric cylinders under consideration [Fig. \ref{fig:cross}(a)]. Noteworthy also is that the LC-symmetry threshold depends on the operating frequency. Calculations show that the closer is the operating frequency to cutoff frequencies of the coupled dielectric cylinders, the larger is the LC-symmetry threshold.

\begin{figure}[t!]
\center
\centering
\fbox{\includegraphics[width=\linewidth]{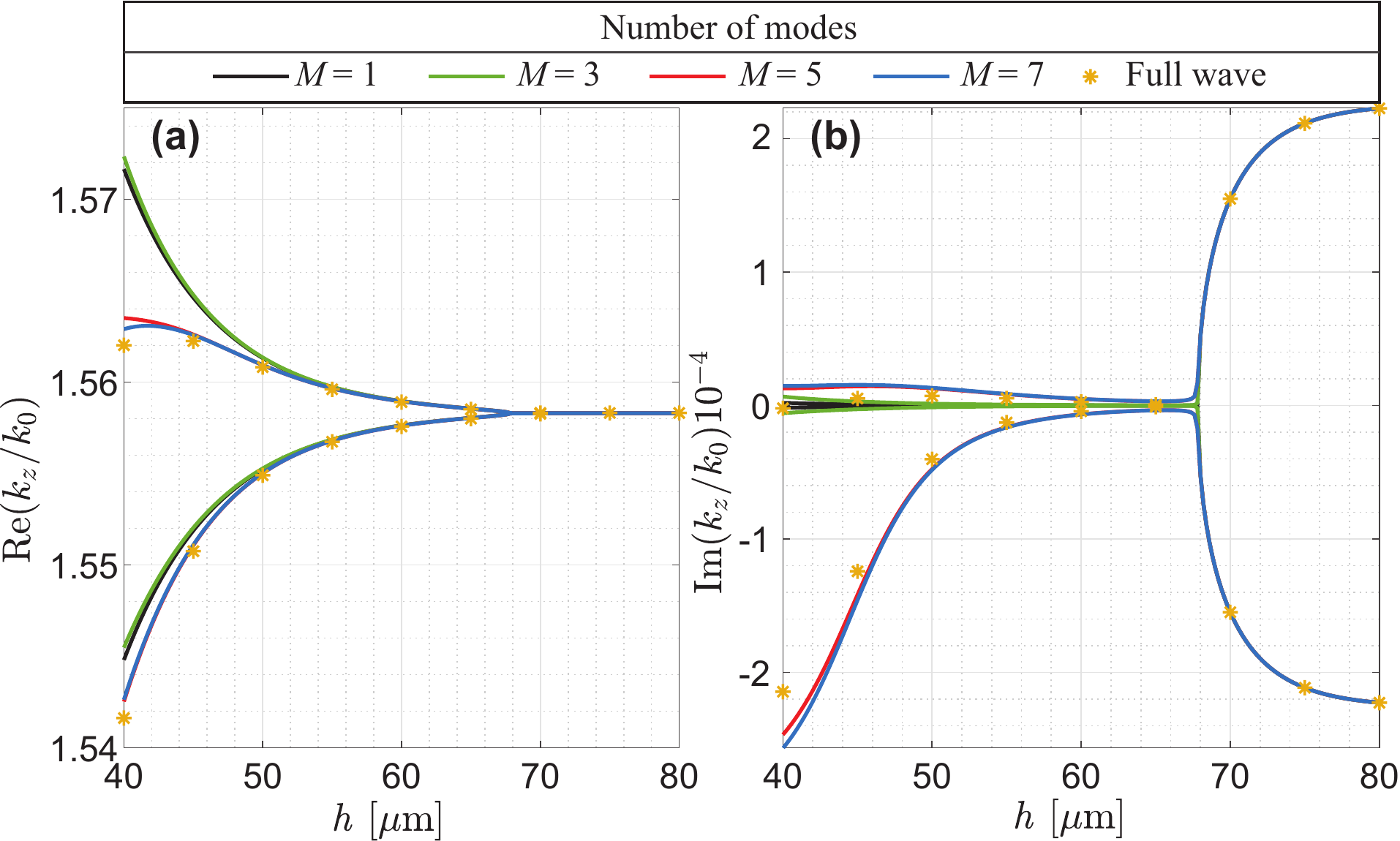}}
\caption{(a) Real and (b) imaginary parts of the axial wavenumber for coupled TM$_{01}$ and TM$_{02}$ modes versus the distance $h$ between dielectric cylinders ($f=5.4961$ [THz]).}
\label{fig:aprox}
\end{figure}

A loss compensation phenomenon for coupled dielectric cylinders is impaired below the LC-symmetry threshold. The reason is the high-order hybrid modes with azimuthal indices $n=\pm1,\pm2,\pm3...~$. This can be seen from Fig. \ref{fig:aprox}, which additionally demonstrates rapid convergence of the developed theoretical approach with the number $M=2N+1$ of basis modes involved in the truncated sums of Eqs.~(\ref{field_1})-(\ref{field_3}). The results obtained are supported by the full-wave simulations with the COMSOL Multiphysics{\textsuperscript{\tiny\textregistered}} software. Compared to these simulations, our approach benefits much less computational time and computer requirements for similar accuracy of numerical results. 

\begin{figure}[h]
\center
\fbox{\includegraphics[width=\linewidth]{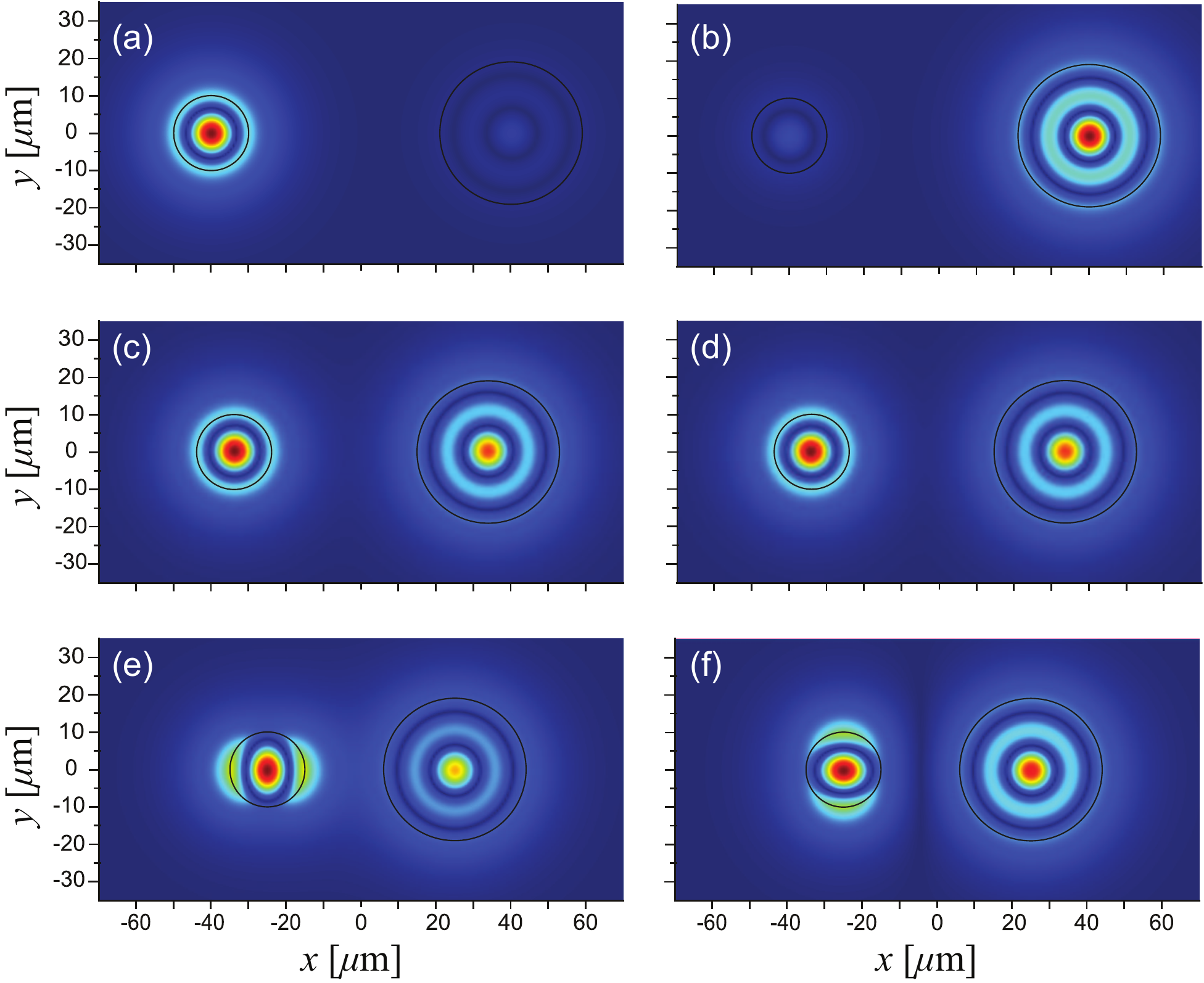}}
\caption{Distribution of $|E_z|$ for the coupled TM$_{01}$ and TM$_{02}$ modes of a pair of dielectric cylinders which are disposed on a distance (a) and (b) $h=80$ $[\mu$m], (c) and (d) $h=67.5$ $[\mu$m], and (e) and (f) $h=50$ $[\mu$m].}
\label{fig:field}
\end{figure}

For the operating frequency $f=5.4961$ [THz], we additionally plot the absolute value of the axial electric field in coupled dielectric cylinders made of gain and lossy materials for several values of the distance $h$ (Fig. \ref{fig:field}). It can be seen that the field of each cylinder oscillates separately above the LC-symmetry threshold [Figs. \ref{fig:field}(a) and \ref{fig:field}(b)]. At this threshold and below, the eigenmodes are in the form of coupled modes of both cylinders [Figs. \ref{fig:field}(c) and \ref{fig:field}(d)]). The effect of high-order axial harmonics manifests itself in asymmetry of the field distribution of the coupled modes [Figs. \ref{fig:field}(e) and \ref{fig:field}(f)]. This effect grows in importance with the decrease in the distance $h$ between dielectric cylinders with gain and loss. The results obtained give a better insight into mechanism of loss compensation in loss-gain systems and can be extended to systems composed of many cylinders with higher-order exceptional points. It is expected that these results can facilitate further development of a new generation of active nanodevices for lasing and optical sources.

To conclude, a rigorous analytical approach has been developed and applied to calculate the LC-symmetry threshold for a pair of loss-gain dielectric cylinders operated in different TM modes. It has been found that this threshold, which is also known as EP, corresponds to full loss compensation in the coupled cylinders and can be achieved for a certain distance between the cylinders, provided that their permittivity and dimensions are properly selected. Above this threshold, the TM modes of different cylinders are weakly coupled to each other, while their dispersion curves exhibit crossing behavior. At the LC-symmetry threshold, these modes coalesce into a single mode with a real eigenvalue. It has been shown that this phenomenon requires the coupled cylinders to have unequal loss and gain tangents. It has been found that below the LC-symmetry threshold the eigenmodes are coupled modes of both cylinders and feature anti-crossing dispersion curves. These modes are hybrid due to the contribution of higher azimuthal harmonics to eigenfields. It has been shown that these harmonics deteriorate the loss compensation phenomenon for small distances between coupled cylinders and lead to the asymmetrical distribution of the eigenfields. The results obtained have been validated against the full-wave simulations by the COMSOL Multiphysics{\textsuperscript{\tiny\textregistered}} software. 

We believe that the above-described phenomenon is of the general nature and can be realized for LC-symmetric waveguide systems operated in coupled modes of any (hybrid) type. 

\medskip
{\bf Funding.} National Key R\&D Program of China (Grant 2018YFE0119900); State Committee on Science and Technology of Belarus (Grant F20KITG-010).

\medskip
{\bf Disclosures.} The authors declare no conflicts of interest.

\bibliography{cylinders}
\bibliographyfullrefs{cylinders}

\end{document}